\documentclass[prd,twocolumn,showpacs,nofootinbib,preprintnumbers,floatfix]{revtex4}  
\usepackage[plainpages=false, colorlinks=true, anchorcolor=blue, linkcolor=blue, citecolor=blue, bookmarks=false]{hyperref}
\usepackage{graphicx}  
\usepackage{dcolumn}   
\usepackage{bm}        
\usepackage{amssymb}   
\usepackage{natbib}
\usepackage{amsmath}   
\usepackage{longtable}  
\usepackage{float} \usepackage[utf8]{inputenc}
\usepackage{tasks}

\begin{document}

\newcommand{\apjl}{Astrophys. J. Lett.}
\newcommand{\apjs}{Astrophys. J. Suppl. Ser.}
\newcommand{\aap}{Astron. \& Astrophys.}
\newcommand{\aj}{Astron. J.}
\newcommand{\pasp}{PASP}
\newcommand{\araa}{Ann. Rev. Astron. Astrophys. } 
\newcommand{\aapr}{Astronomy and Astrophysics Review}
\newcommand{\ssr}{Space Science Reviews}
\newcommand{\mnras}{Mon. Not. R. Astron. Soc.}
\newcommand{\apss} {Astrophys. and Space Science}
\newcommand{\jcap}{JCAP}
\newcommand{\na}{New Astronomy}
\newcommand{\pasj}{PASJ}
\newcommand{\pasa}{Pub. Astro. Soc. Aust.}
\newcommand{\physrep}{Physics Reports}
\newcommand{\rthis}[1]{\textcolor{black}{#1}}

\title{A test of the evolution of gas  depletion factor in galaxy clusters using strong gravitational lensing systems}

\author{R. F. L. Holanda$^{1,2,3}$}\email{holandarfl@fisica.ufrn.br}

\author{Kamal Bora$^{4}$}\email{ph18resch11003@iith.ac.in}

\author{Shantanu Desai$^{4}$}\email{shntn05@gmail.com}

\affiliation{$^1$Departamento de F\'{\i}sica, Universidade Federal do Rio Grande do Norte,Natal - Rio Grande do Norte, 59072-970, Brasil}

\affiliation{$^2$Departamento de F\'{\i}sica, Universidade Federal de Campina Grande, 58429-900, Campina Grande - PB, Brasil}

\affiliation{$^3$Departamento de F\'{\i}sica, Universidade Federal de Sergipe, 49100-000, Aracaju - SE, Brazil}

\affiliation{$^4$ Department of Physics, Indian Institute of Technology, Hyderabad, Kandi, Telangana-502285, India }

\begin{abstract}
  In this \rthis{work}, we discuss  a new method to probe the redshift evolution of the  gas depletion factor, i.e. the ratio by which the gas mass fraction of galaxy clusters is depleted with respect to the  universal  mean of baryon fraction. The dataset we use for this purpose consists of 40  gas mass fraction measurements measured at $r_{2500}$   using  Chandra  X-ray observations, strong gravitational lensing sub-samples  obtained from SLOAN Lens ACS + BOSS Emission-line Lens Survey (BELLS) + Strong Legacy Survey SL2S + SLACS. For our analysis,  the validity of the cosmic distance duality relation is assumed. We find a mildly decreasing trend  for the gas depletion factor as a function of redshift at about 2.7$\sigma$.  
\end{abstract}

\maketitle

\section{Introduction}

Galaxy clusters are  the largest gravitationally bound
structures in the Universe, and hence are  a powerful probe of the evolution and structure formation in  the Universe at redshifts $z < 2$ (see \cite{Mantz:2014xba,Vikhlininrev} for  detailed reviews).  They are also wonderful laboratories for fundamental Physics measurements~\cite{Carbone,Desai18}
One can obtain constraints on the cosmological parameters,  if we assume that the X-ray gas mass
fraction, $f_{gas}$, of hot, massive and relaxed galaxy clusters does not evolve with redshift \cite{1996PASJ...48L.119S,Allen2007,Ettori2009,Mantz:2014xba,pen97,white93}.  In order to quantify the baryon content  and its possible time evolution, crucial to cosmological tests with  gas mass fraction measurements, hydrodynamic simulations have been used to calibrate the gas depletion factor, $g(z)$, viz. the ratio by which the gas mass fraction of galaxy clusters is depleted with respect to the  universal  mean of the baryon fraction. \citet{2013MNRAS.431.1487P} and \citet{2013ApJ...777..123B},  have previously shown using simulations  of hot, massive, and dynamically relaxed galaxy clusters ($M_{500}>10^{14} M_{\odot}$) involving   different physical processes, that by modelling  a possible time evolution for $g(z)$, given by $g(z) \equiv g_0(1+g_1 z)$ \footnote{Note that elsewhere in literature (eg.~\cite{Bora2021EPJC}),the parametric form for the  gas depletion factor is usually denoted by  $ \gamma(z) = \gamma_0(1+\gamma_1 z)$},    one  obtains: $0.55 \leq g_0 \leq 0.79$ and $-0.04 \leq g_1 \leq 0.07$, depending on the physical processes that are included in the simulations (see Table 3 in \citep{2013MNRAS.431.1487P} for values obtained at $r_{2500}$). Therefore, no significant trend of $g(z)$ as a function of redshift was found in their simulations. For their analysis, ~\citet{2013MNRAS.431.1487P} considered $f_{gas}$ as a cumulative quantity up to $r_{2500}$, the radii at which the mean
cluster density is 2500 times the critical density of the
Universe at the cluster’s redshift. Recently, by using the cosmo-OverWhelmingly Large Simulation,  the Ref. \cite{Brun2017MNRAS.466.4442L} showed that in most realistic models of intra-cluster medium, which include active galactic nuclei feedback, the slopes of the various mass-observable relations deviate substantially from the self-similar one. Self-similarity implies that  galaxy clusters  are identical objects when scaled by their mass~\cite{Kaiser86},  or in other words, galaxy clusters form via a single gravitational collapse, particularly at late times and for low-mass clusters. A constant non-varying gas depletion factor is also expected from self-similarity~\cite{Mohr99}.

On the other hand, there have been a number of works, which   have estimated the depletion factor directly from observations, without using simulations. For instance, the Ref.\cite{Holanda2017JCAP}  carried out an analysis using 40 $f_{gas}$ measurements  observed by the Chandra X-ray telescope  provided by the Ref. \cite{Mantz:2014xba}  along with Type Ia supernovae observations,  with  priors  on $\Omega_b$ and $\Omega_M$ from  the Planck  results \cite{Planck2015}, and assuming the validity of the cosmic distance duality relation (CDDR) \cite{2007GReGr..39.1047E}. No evolution of  the gas depletion factor with redshift was found from this aforementioned analysis.
In a follow-up work, Ref. \cite{Holanda2018} used a completely independent set of 38 Chandra X-ray  $f_{gas}$ measurements  in the redshift range $0.14 \leqslant z \leqslant 0.89$  provided by the Ref. \cite{Laroque2006ApJ}, along with angular diameter distances from X-ray/SZ measurements and priors from Planck results \cite{Planck2015}. Unlike~\cite{Holanda2017JCAP}, they did not use CDDR to derive the angular diameter distance. The aforementioned work reported $g_0$ =  $0.76\pm0.14$ and $g_1$ = $-0.42_{-0.40}^{+0.42}$, which also implies no redshift evolution for the  gas depletion factor. The gas mass fractions calculated by the Ref.~\cite{Laroque2006ApJ} were obtained within $r_{2500}$, while those from the Ref.~\cite{Mantz:2014xba}  were calculated in spherical shells at radii near $r_{2500}$.

Constraints on the evolution of $g(z)$ using gas mass fraction measurements within $r_{500}$ have also been obtained.
\citet{Zheng2019EPJC} used 182 galaxy clusters detected by the Atacama Cosmology Telescope Polarization experiment~\cite{Hilton18} and  cosmic chronometers~\cite{Jimenez02}, and found a mild decrease of $g(z)$ as a function of redshift ( $g(z)$ is smaller at high z).
Most recently, \citet{Bora2021EPJC} also looked for a time evolution of $g(z)$ using gas mass fraction measurements at $r_{500}$,  from  both the SPT-SZ \cite{2018MNRAS.478.3072C} and Planck Early Sunyaev-Zeldovich effect cluster data \cite{Planck2011A&A...536A...8P} along with cosmic chronometers and priors from Planck results \cite{Planck2020A&A}. Conflicting  results between both the samples were found: for SPT-SZ, $g(z)$ was found to be decreasing as a function of redshift (at more than 5$\sigma$), whereas a positive trend with redshift was found for Planck ESZ data (at more than 4$\sigma$). Their results implied that  one cannot use  $f_{gas}$ values  at $r_{500}$ as a stand-alone probe  for any model-independent cosmological tests. 

In this letter, we propose a new observational test for   the  time evolution of the gas depletion factor by using  multiple large scale structure probes, namely, galaxy cluster gas mass fraction measurements~\cite{Mantz:2014xba} along with  strong gravitational lenses (SGL) observations obtained from SLOAN Lens ACS + BOSS Emission-line Lens Survey (BELLS) + Strong Legacy Survey SL2S + SLACS \cite{Leaf2018}. For our analysis, we assume the validity of the CDDR \cite{2007GReGr..39.1047E}. By considering a possible time evolution for $g(z)$, such as $g(z) = g_0(1+g_1 z)$, and analyses by using three SGL sub-samples  separately (segregated according to  their mass intervals), we obtain an error-weighted average given by:  $g_1 = -0.15 \pm 0.055$ at $1\sigma$ c.l.
Therefore, we infer  a mild time evolution at about $2.7\sigma$ from our analysis.
We note that unlike previous works in this area, our results are independent  of the  Planck  cosmological parameters ($\Omega_b$ and $\Omega_M$). However, a flat universe model with $\Omega_{tot}=1$ is still  assumed.

This letter is organized as follows. Section~\ref{methodology} explains the methodology adopted in this work. In section~\ref{data}, we present the data sample used for our analysis. Section~\ref{sec:analysis} describes our analysis and results. Our conclusions are presented in Section~\ref{sec:conclusions}. 

\section{Methodology}
\label{methodology}

In this section, we discuss some aspects of SGL systems and gas mass fractions, and show how one can combine these observations in order to put constraints on the gas depletion factor.

\subsection{Strong Gravitational Lensing Systems}

 Strong gravitational lensing has proved to be a powerful diagnostic  of modified gravity theories  and cosmological models, as well as  fundamental physics~\cite{1992grle.book.....S,2010CQGra..27w3001B}. \rthis{These systems  have been used to constrain the PPN $\gamma$ parameter, cosmic curvature~\cite{Kumar21,Wei22,Wang20,Cao17}, variations of speed of light~\cite{Liu21}, variations of fine structure constant~\cite{Colaco21}, Hubble constant~\cite{Qi21}, tests of FLRW metric~\cite{Caof}, etc.}
  Usually,  a strong lens system  consists of  a foreground galaxy or  a cluster of galaxies positioned between a source (quasar) and an observer, where the multiple-image separation from the source depends only on the lens and the source angular diameter distance (see, for instance,  Refs.~\cite{Futamase,Marek,Grillo,Cao12,Cao2015ApJ,Holanda:2016msr,Cao2018,Amante:2019xao,Lizardo:2020wxw,boraDMevolution}, where SGL systems were used recently as a cosmological tool). By using  the simplest model assumption (the  singular isothermal sphere) to describe the SGL systems,  the Einstein radius ($\theta_E$), a fundamental quantity,  can be defined as \cite{2010CQGra..27w3001B,Cao2015ApJ}:

\begin{equation}
\theta_E = 4\pi \frac{D_{A_{ls}}}{D_{A_{s}}} \frac{\sigma_{SIS}^{2}}{c^2}
\label{eq:thetaE_SIS}
\end{equation}
In this equation,  $D_{A_{ls}}$ is the angular diameter distance from the lens to the source; $D_{A_{s}}$ denotes the angular diameter distance from the observer to the source; $\sigma_{SIS}$ is the velocity dispersion caused by the lens mass distribution; and $c$ is the speed of light.

For  our analyses, a  flat universe is considered and the following  observational quantity derived  for SGL systems is used~\cite{Liao19}: 
\begin{equation}
\label{eq:D_SIS}
D={\frac{D_{A_{ls}}} {D_{A_s}}}=\frac{{\theta}_E c^2}{4{\pi} \sigma^2_{SIS}}
\end{equation}
For a flat universe, the comoving distance $r_{ls}$ is given  by
$r_{ls}=r_s-r_l$~\cite{2010CQGra..27w3001B}, and using $r_s=(1+z_s)D_{A_s}$, $r_l=(1+z_l)D_{A_l}$  and $r_{ls}=(1+z_s)D_{A_{ls}}$, we obtain the following:
\begin{equation}
\label{eq:D}
D= 1 - \frac{(1+z_l)D_{A_{l}}}{(1+z_s)D_{A_{s}}}
\end{equation}
Finally, by the validity of the CDDR relation
 $D_L=(1+z)^2D_A$~\cite{2007GReGr..39.1047E}, Eq.~\ref{eq:D} can be re-cast as follows:
\begin{equation}
\label{eq:final_D}
\frac{(1+z_s)}{(1+z_l)}= (1-D)\frac{D_{L_s}}{D_{L_l}}
\end{equation}

\subsection{Gas mass fraction}

The cosmic gas mass fraction can be  defined as $f_{gas} \equiv \Omega_b/\Omega_M$ (where $\Omega_b$ and $\Omega_M$ are the baryonic and total matter density parameters, respectively), and the constancy  of this quantity within massive, relaxed clusters within $r_{2500}$ has been used to constrain cosmological parameters by using the following equation (see, for instance, \cite{Mantz:2014xba}) 
\begin{equation}
\label{fgas}
f_{gas}(z) = A(z) K g(z)  \left[\frac{\Omega_b}{\Omega_{M}}\right] \left(\frac{D_L^*}{D_L}\right)^{3/2}\
\end{equation}
Here,  the observations are done in the X-ray band. The asterisk in Eq.~\ref{fgas} denotes the corresponding quantities for the fiducial model used in the observations to obtain  $f_{gas}$ (usually a flat $\Lambda$CDM model with Hubble constant $H_0=70$ km s$^{-1}$ Mpc$^{-1}$ and the present-day total matter density parameter $\Omega_M=0.3$), $A(z)$ represents the angular correction factor, which is very close to unity in almost all the cases, and hence can be neglected. The $K$ factor  is the instrument calibration constant, which also accounts for any bias in the  mass due to non-thermal pressure and bulk motions in the  baryonic gas~\cite{Allen2007,Mantz:2014xba,2013MNRAS.431.1487P,2013ApJ...777..123B}.  Twelve galaxy clusters used in this present work are also part of the ``Weighing the Giants'' sample \cite{2016MNRAS.457.1522A}, and this work found the $K$ factor to be constant, i.e.,  no significant trends with
mass, redshift, or any  morphological indicators were found. Therefore, we also posit the same in  this work. The quantity $g(z)$, which is of interest here is again modelled in the same way as previous works, i.e. $g(z)=g_0(1+g_1z)$. The ratio in the parenthesis of Eq.~\ref{fgas} encapsulates the expected variation in $f_{gas}$ when the underlying cosmology is varied, which makes the analyses with gas mass fraction measurements model-independent. Finally, it is important to stress that  Eq.~\ref{fgas} is obtained only after assuming the  validity of CDDR (see Ref.~\cite{Gon2012} for details). 
In the last decade, a plethora of analyses using myriad observational data have been undertaken in order to establish whether or
not the CDDR holds in practice.   A succinct summary of the latest observational constraints on the CDDR
can be found in Refs.\cite{2021arXiv210401614H,BoraCDDR}, which demonstrate the validity of CDDR  to within 2$\sigma$.

The key equation to our method can be  obtained from combining Eq.~\ref{eq:final_D} and \ref{fgas}, and  by incorporating a possible redshift dependence for the gas depletion factor, such as $g(z)=g_0(1+g_1z)$. In this way, we now obtain:

\begin{equation}
\label{main_equation}
\left[\frac{(1+g_1 z_s)}{(1+g_1 z_l)}\right] = \left[\frac{f_{gas}(z_s)}{f_{gas}(z_l)}\right] \left[\frac{(1+z_s)D^*_{L_l}}{(1+z_l)D^*_{L_s}}\right]^{3/2}(1-D)^{-3/2}
\end{equation}
 As we can see, unlike Ref.\cite{Holanda:2017cmc,Bora2021EPJC}, our results are independent of the $K$ factor, $g_0$, and   $\Omega_b/\Omega_M$.

\begin{figure*}
    \centering
    \includegraphics[width=8.8cm, height=5cm]{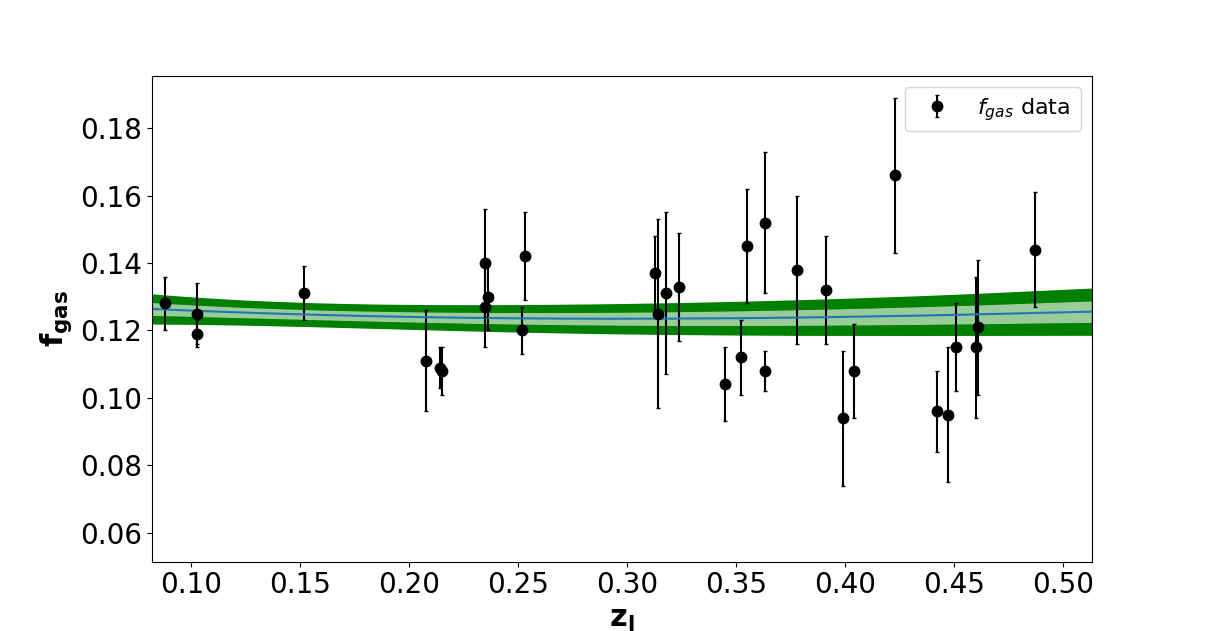}
    \includegraphics[width=8.8cm, height=5cm]{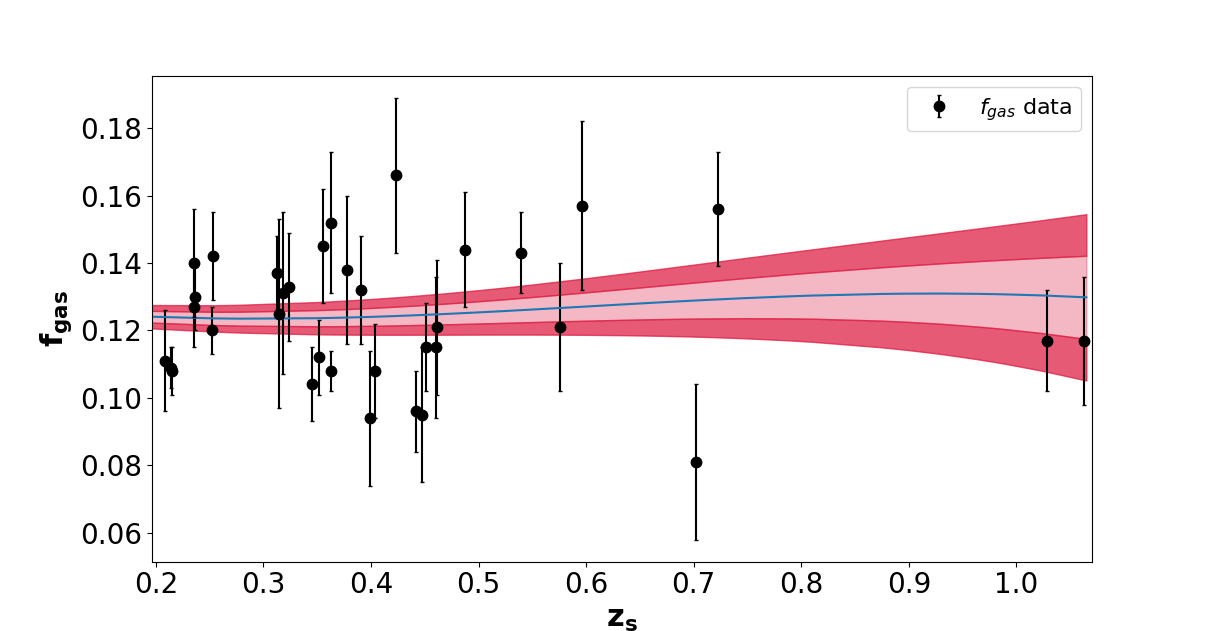} 

    \caption{ (Left) Reconstruction of the gas mass fraction at $z_l$ for the Intermediate Mass range sample using Gaussian Process Regression. The green shaded area indicates 1$\sigma$ and 2 $\sigma$ error bands. (Right) Reconstruction of the gas mass fraction at $z_s$ for the Intermediate Mass range sample using Gaussian Process Regression. The crimson shaded area indicates 1 $\sigma$ and 2 $\sigma$ error bands.  }
    \label{fig:GP}
\end{figure*}

\section{Cosmological data}
\label{data}
\begin{figure}[t]
    \centering
    \includegraphics[width=9.8cm, height=6.8cm]{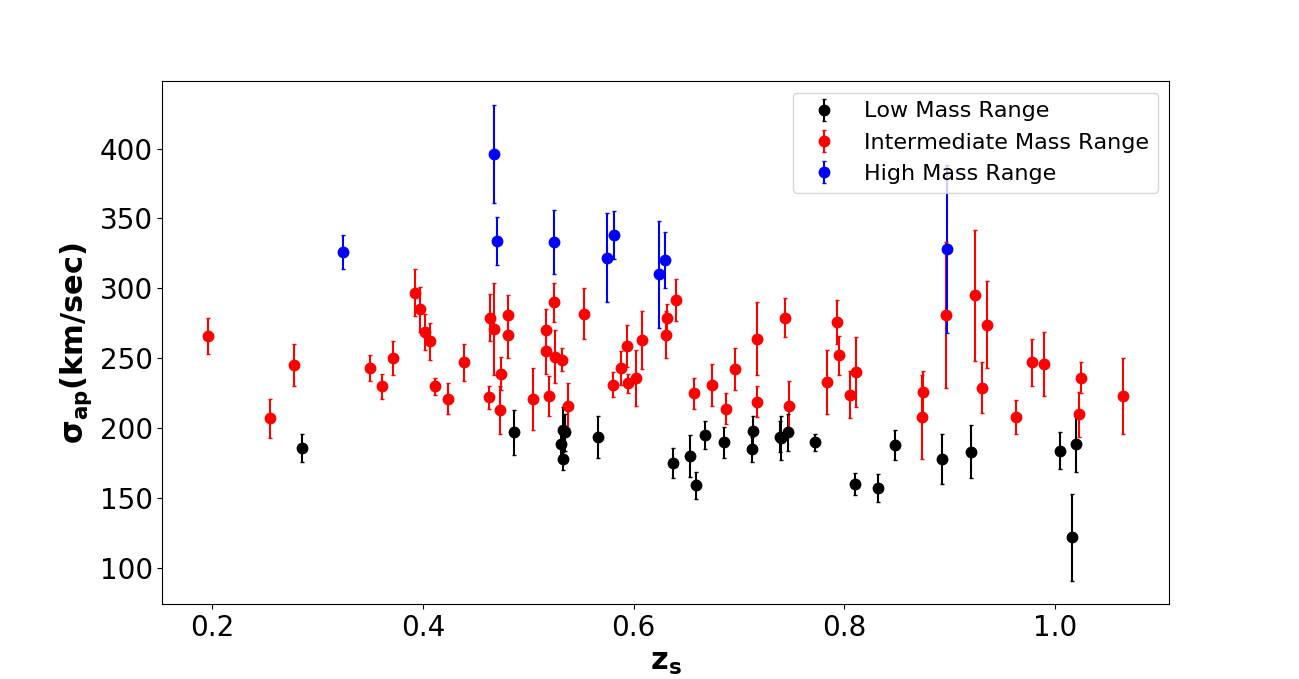} 
    \caption{ The different redshifts range covered by Low, Intermediate, and High mass range SGL sample.  }
    \label{fig:vel_dispersion}
\end{figure}

The data used in our analysis were as follows:
\begin{itemize}
\item 40 X-ray gas mass fraction measurements of hot ($kT \geq  5$ keV), massive and morphologically relaxed  galaxy clusters from  archival Chandra data.~\footnote{While this work was in preparation, there was an update to this dataset~\cite{Mantz22}. For this work we use the 2014 dataset. We shall use the updated dataset in a future work.} \rthis{The exposure times range from 7-350 ksec. (cf Table 1 of ~\cite{Mantz:2014xba}). The morphological diagnostics used to identify relaxed clusters are discussed in ~\cite{Mantz15}.} The galaxy cluster redshift range is $0.078 \leq z \leq 1.063$~\cite{Mantz:2014xba} \rthis{  The restriction to relaxed systems minimizes the systematic biases due to departures from hydrostatic equilibrium and substructure, as well as the scatter due
to these effects, asphericity, and projection.   The bias in the mass measurements from X-ray data arising by assuming hydrostatic equilibrium was calibrated using robust mass estimates for the target clusters from weak gravitational lensing \cite{2016MNRAS.457.1522A}, reducing systematic uncertainties.  Furthermore, these authors measured the gas mass fraction in spherical shells at radii near at $r_{2500}$ ($\approx 1/4$ of virial radius), instead of using  the cumulative fraction integrated over all radii ($< r_{2500}$) as in several previous works present in literature.  The data were  fitted with a non-parametric model for the deprojected, spherically symmetric intracluster medium density and temperature profiles. In this model, the cluster atmosphere is described as a set of concentric,
spherical shells (near at $r_{2500}$), and, within each shell, the gas is assumed to be isothermal (see more details in \cite{2016MNRAS.456.4020M}). Therefore the gas mass fraction is agnostic to details of the temperature and gas density profile of the sort discussed in ~\cite{Cao16,CaoZhu11}, which would mainly affect   gas mass fraction measurements only at $R_{500}$.
As pointed by these authors, the observations were restricted  to the most self-similar and accurately measured regions of clusters, significantly reducing systematic uncertainties compared to previous work. The dark matter  profile of the clusters were modeled by the so-known Navarro-Frank-White form.}
\item We also consider sub-samples  from a specific catalog containing 158 confirmed sources of strong gravitational lensing systems \cite{Chen2019,Leaf2018}. This  compilation includes 118 SGL systems identical to the compilation of \cite{Cao2015ApJ}, which was obtained from a combination of SLOAN Lens ACS, BOSS Emission-line Lens Survey (BELLS), and Strong Legacy Survey SL2S, along with 40 additional systems recently discovered by SLACS and pre-selected by \cite{2017ApJ...851...48S} (see Table I in \cite{Leaf2018}).  Several studies have shown that the slopes of  the density profiles for individual galaxies show a non-negligible deviation from the Singular isothermal sphere profile~\cite{Koopmans2009,Auger2010,Barnab2011,Sonnenfeld2013,Cao2016,HolandaSaulo,Chen2019}. Therefore, for the mass distribution of lensing systems, the  power-law model is assumed. This model considers  a spherically symmetric mass distribution with a more general power-law index $\gamma $, namely $\rho \propto r^{-\gamma}$. In this approach $\theta_E$ is given by:
\begin{equation}
\theta_E =   4 \pi
\frac{\sigma_{ap}^2}{c^2} \frac{D_{ls}}{D_s} \left[
\frac{\theta_E}{\theta_{ap}} \right]^{2-\gamma} f(\gamma),
\label{thetaE}
\end{equation}
where $\sigma_{ap}$ is the  stellar velocity dispersion inside an aperture of size $\theta_{ap}$ and
\begin{eqnarray} \label{f factor}
f(\gamma) &=& - \frac{1}{\sqrt{\pi}} \frac{(5-2 \gamma)(1-\gamma)}{(3-\gamma)} \frac{\Gamma(\gamma - 1)}{\Gamma(\gamma - 3/2)}\nonumber\\
          &\times & \left[ \frac{\Gamma(\gamma/2 - 1/2)}{\Gamma(\gamma / 2)} \right]^2
\end{eqnarray}
Thus, we obtain: 
\begin{equation} 
\label{NewObservable}
 D=\frac{D_{A_{ls}}}{D_{A_{s}}} = \frac{c^2 \theta_E }{4 \pi \sigma_{ap}^2} \left[ \frac{\theta_{ap}}{\theta_E} \right]^{2-\gamma} f^{-1}(\gamma)
\end{equation}
For $\gamma = 2$,  the singular isothermal spherical distribution is recovered. All the  terms necessary  to evaluate $D$ can be found in Table 1 of \cite{Leaf2018}. However, the complete dataset (158 points) is culled to  98 points, after the following cuts: $z < 1.061$ and  $D \pm \sigma_D <1 $ ($D > 1$ represents a non-physical region). The compilation considered here contains only those SGL systems with early type galaxies acting as lenses. Each data point contains estimated apparent and Einstein radius, with spectroscopically measured stellar apparent velocity dispersion, as well as both the lens and the source redshifts.

We should point out that  some works have recently explored a possible time evolution of the mass density power-law index \cite{Cao2016,HolandaSaulo,Amante:2019xao,Chen2019}. No significant evolution has been found in these works. On the other hand, these results indicate that it is prudent to  use low, intermediate, and high-mass galaxies separately in any cosmological analyses. As commented in \cite{Cao2016}, elliptical galaxies with velocity dispersions smaller than $200$ km/s may be classified roughly as relatively low-mass galaxies, while those with velocity dispersion larger than $300$ km/s may be treated as relatively high-mass galaxies. Naturally, elliptical galaxies with velocity dispersion between $200-300$ km/s may be classified as intermediate-mass galaxies. Therefore, in our analyses, we work with three distinct sub-samples consisting of 26, 63, and 9 data points with low, intermediate, and high velocity dispersions, respectively.  The redshift range covered by these three samples is shown in Figure~\ref{fig:vel_dispersion}. As we can see, all the three samples cover roughly the same redshift range.

To constrain $g_1$ by using  Eq.~\ref{main_equation},   gas mass fraction measurements  at the  lens and source redshifts (for each SGL system) are required. These quantities are obtained by applying Gaussian Process Regression (GPR)~\cite{seikel12,Haveesh} using the 40 gas mass fraction measurements compiled by  Ref.\cite{Mantz:2014xba} in order to estimate the gas mass fraction at any arbitrary redshift.  In  Fig.~\ref{fig:GP}, we show  the result of reconstruction of the gas mass fraction at $z_l$ and $z_s$ from the Intermediate Mass Range SGL sub-sample for the purpose of illustration. A discussion of GPR and comparison with other non-parametric regression techniques such as ANN can be found in the Appendix. The final results will  also be  sensitive to the choice of the reconstruction technique.
\end{itemize}
\vspace{0.2cm}

\begin{table*}[]
\caption{\label{tab:table1}. Constraints on the parameters $\gamma$ and $g_1$ for different $\sigma_{ap}$ range used in this analysis as discussed in Sect.~\ref{sec:analysis}.}
    \centering
    \begin{tabular}{|l|c|c|c|r|} \hline
    \textbf{Sample} & \boldmath$\sigma_{ap}$\textbf{(km/sec)} & \boldmath$\gamma$ & \boldmath$g_1$\\ \hline 
      Low\ & $\sigma_{ap}< 200$ & $1.913^{+0.009}_{-0.010}$  &  $0.172^{+0.173}_{-0.167}$ \\
      Intermediate & $200 < \sigma_{ap}< 300$ & $2.041\pm0.012$ & $-0.188^{+0.060}_{-0.059}$ \\
      High &  $\sigma_{ap}> 300$  & $1.983^{+0.082}_{-0.081}$ & $-0.137^{+0.319}_{-0.254}$\\
      
      \hline 
      
    \end{tabular}

\end{table*}
\section{Analysis and Results} 
\label{sec:analysis}

The joint constraints on the power law index, $\gamma$ and redshift dependent part of the gas depletion factor, $g_1$  can be obtained by maximizing the likelihood distribution function, ${\cal{L}}$  given by

\begin{widetext}
\begin{equation}
    \label{eq:logL}
   -2\ln\mathcal{L} = \sum_{i=1}^{n} \ln 2\pi{\sigma_{i}^2}+ \sum_{i=1}^{n}\frac{\left(\Phi(g_1,z_i)- \left[\frac{f_{gas}(z_s)}{f_{gas}(z_l)}\right] \left[\frac{(1+z_s)D^*_{L_l}}{(1+z_l)D^*_{L_s}}\right]^{3/2}(1-D)^{-3/2}\right)^2}{\sigma_{i}^2} , 
\end{equation}          \end{widetext}

where 
\begin{equation}
  \Phi(g_1,z_i) =  \left[\frac{(1+g_1 z_s)}{(1+g_1 z_l)}\right]
\end{equation}

\begin{figure*}
    \centering
    \includegraphics[width=8.9cm, height=6.5cm]{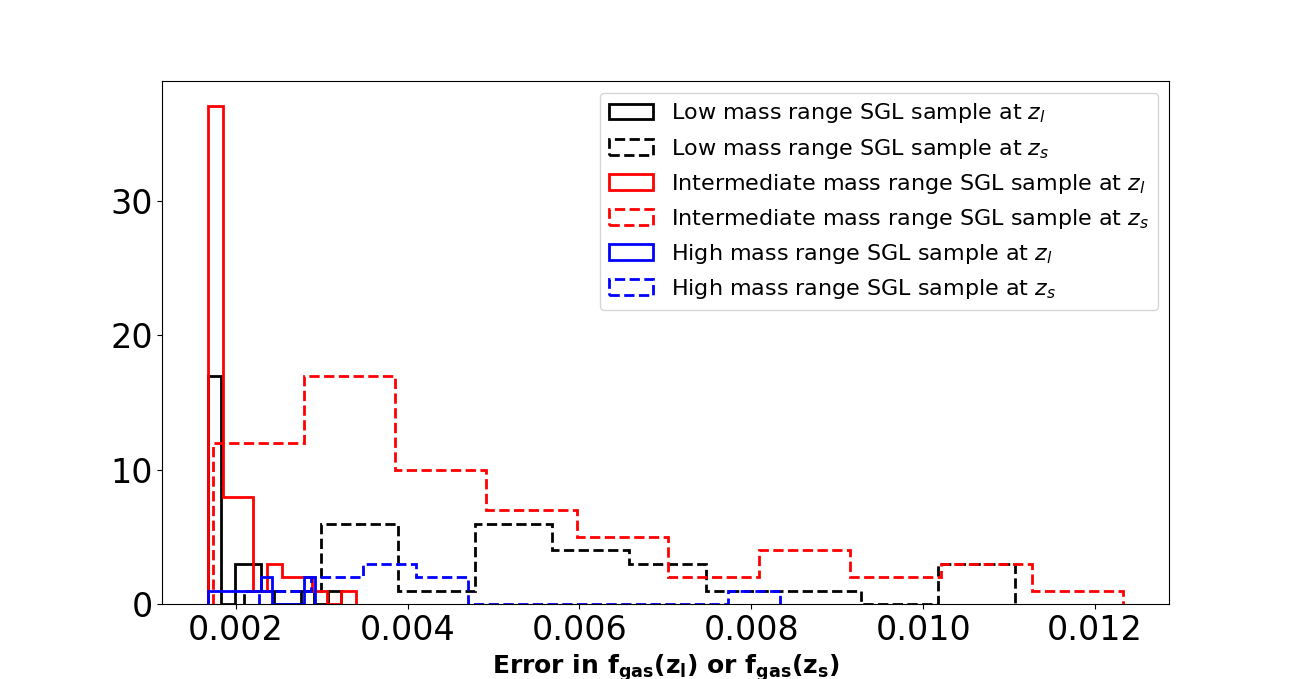}
    \includegraphics[width=8.9cm, height=6.5cm]{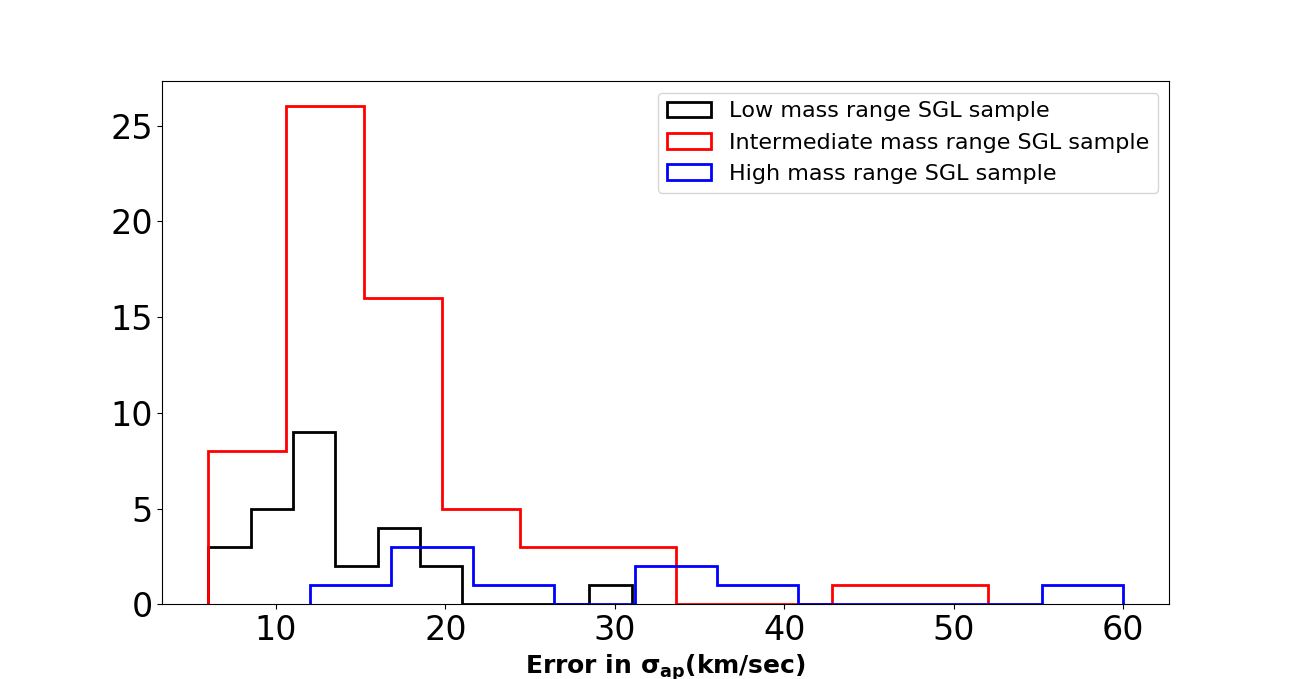} 

    \caption{(Left) The distribution of error in $f_{gas}$ for Low, Intermediate, and High mass range sample at $z_l$ and $z_s$.  (Right) The distribution of error in $\sigma_{ap}$ for Low, Intermediate, and High mass range SGL sample.  }
    \label{fig:histogram}
\end{figure*}

Here, $\sigma_i$   denotes the statistical errors associated with the gravitational lensing observations  (see Table A1 from \cite{Chen2019}) and gas mass fraction measurements (see table II in \cite{Mantz:2014xba}), and are obtained by using standard propagation errors techniques. We have assumed that the errors in the gas mass fraction measurements and lensing observations are uncorrelated, as they correspond to to different systems.  The distribution of errors in $f_{gas}$ at $z_l$ and $z_s$ can be found in the left panel of  Figure~\ref{fig:histogram}, whereas  the same  for errors in the velocity dispersion, $\sigma_{ap}$ can be found in the right panel of  Figure~\ref{fig:histogram}. We can see that the errors are roughly comparable for all the data points, so that no one system will dominate the results.
Note that the quantity $D$ that appears in the log-likelihood (Eq.~\ref{eq:logL})  depends on the parameter $\gamma$ via Eq.~\ref{NewObservable}.  The value of $n$ is equal to 26, 63, 9  for the low, intermediate, and high velocity dispersions, respectively.  
We maximize our likelihood function using the  $\tt{emcee}$ MCMC sampler~\cite{emcee} in order to estimate the free parameters used in Eq.~\ref{eq:logL}, viz. $\gamma$ and $g_1$. 

\begin{figure}[t]
    \centering
    \includegraphics[width=10cm, height=8cm]{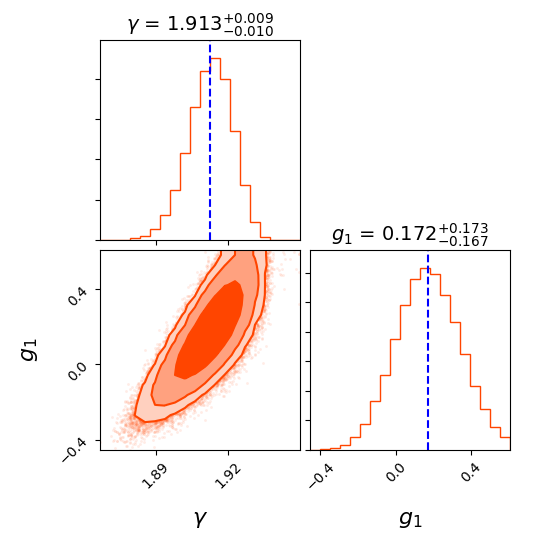} 
    \caption{\textbf{For Low mass range SGL sample:} The 1D marginalized likelihood distributions along with 2D marginalized constraints showing the 68\%, 95\%, and 99\% credible regions for the parameters $\gamma$ (from the power law model describing the SGL systems) and $g_1$ (from the gas depletion factor $g(z)$), obtained using the {\tt Corner} python module~\cite{corner}.}
   \label{fig:low}
    
\end{figure}

\begin{figure}[t]
    \centering
    \includegraphics[width=10cm, height=8cm]{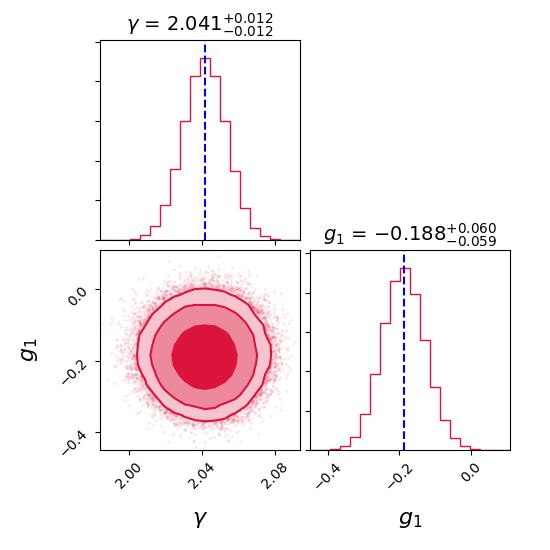} 
    \caption{\textbf{For Intermediate mass range SGL sample:} The 1D marginalized likelihood distributions along with 2D marginalized constraints showing the 68\%, 95\%, and 99\% credible regions for the parameters $\gamma$ (from the power law model describing the SGL systems)  and $g_1$ (from the gas depletion factor $g(z)$), obtained using the {\tt Corner} python module~\cite{corner}.}
    \label{fig:intermediate}
\end{figure}

\begin{figure}[t]
    \centering
    \includegraphics[width=10cm, height=8cm]{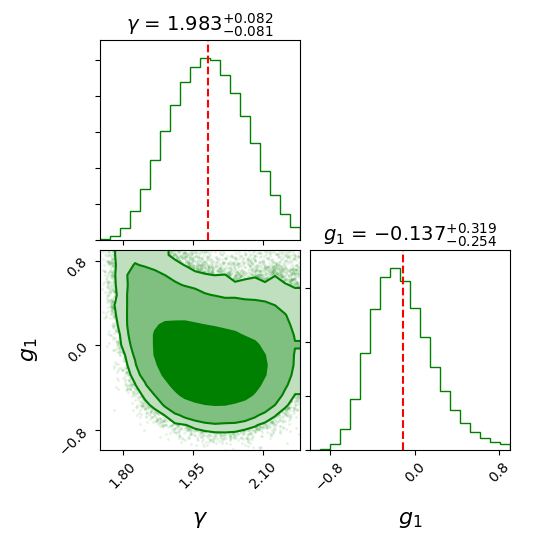} 
    \caption{\textbf{For High mass range SGL sample:} The 1D marginalized likelihood distributions along with 2D marginalized constraints showing the 68\%, 95\%, and 99\% credible regions for the parameters $\gamma$  (from the power law model describing  the  SGL  systems) and $g_1$ (from the gas depletion factor $g(z)$), obtained using the {\tt Corner} python module~\cite{corner}.}
    \label{fig:high}
\end{figure}

\begin{figure*}[t]
    \centering
    \includegraphics[width=9.8cm, height=6.8cm]{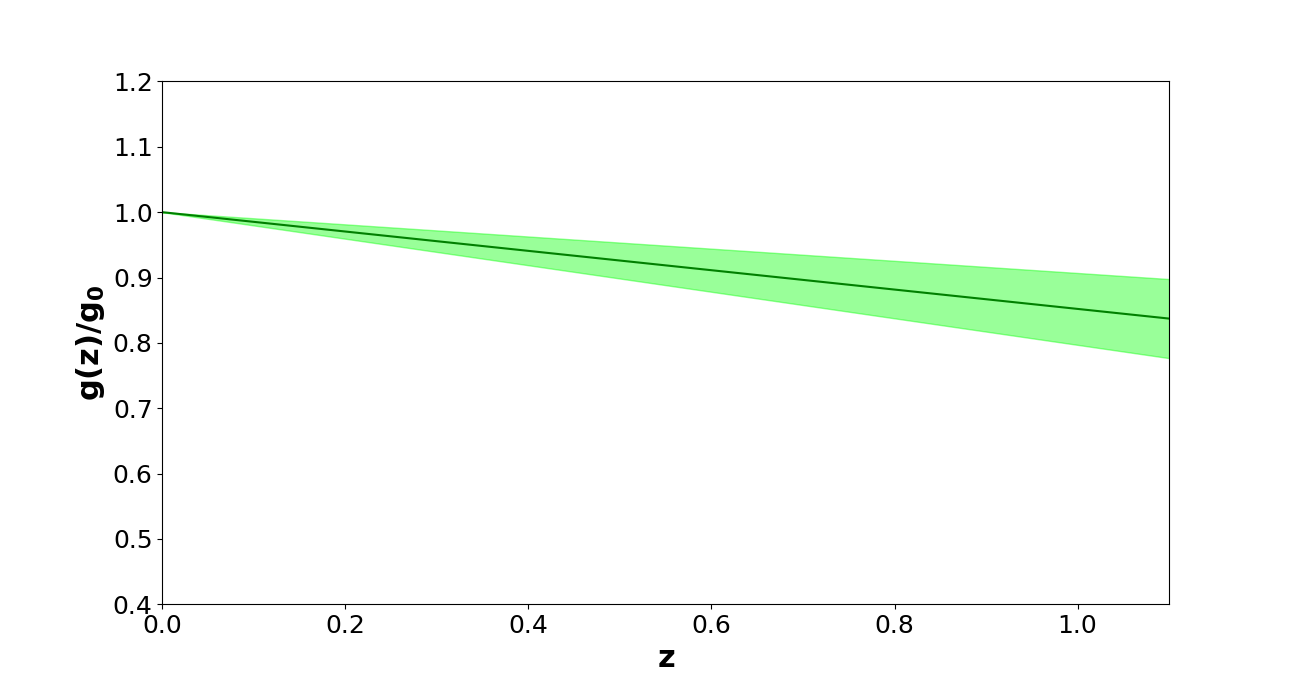} 
    \caption{The evolution of the normalized gas depletion factor along with $1\sigma$ error band as a function of redshift using 40 X-ray $f_{gas}$ measurements near $r_{2500}$ from ~\cite{Mantz:2014xba}.  This figure shows a mild evolution of $g(z)$ with respect to  redshift.}
    \label{fig:evolution}
\end{figure*}

The one-dimensional marginalized posteriors for each parameter along with the 68\%, 95\%, and 99\% 2-D marginalized credible intervals, are shown in Fig.~\ref{fig:low}, Fig.~\ref{fig:intermediate}, and Fig.~\ref{fig:high} for the Low, Intermediate, and High samples, respectively. As we can see, the analyses with the low and high mass SGL sub-samples are in full agreement with no evolution for the depletion factor, while  the intermediate sub-sample shows a non-negligible evolution of $g(z)$ (see Table~\ref{tab:table1}). 
We calculate an error-weighted average using the three estimates  and obtained: $g_1 = -0.15 \pm 0.055$ at  1$\sigma$, showing a mild evolution for  $g(z)$ at 2.7$\sigma$. (see Fig.~\ref{fig:evolution}). 

 We should point out that although a non-constant gas depletion factor has previously been obtained using measurements within $r_{500}$~\cite{Zheng2019EPJC,Bora2021EPJC}, this is the first result  which  finds  a time-varying depletion factor, using gas mass fraction measurements in spherical shells at radii near $r_{2500}$. Previous works \cite{2013MNRAS.431.1487P,2013ApJ...777..123B} using cosmological  hydrodynamic simulations    did not find a significant trend of  $g(z)$ as a function of redshift   for gas mass fraction measurements within $r_{2500}$ or $r_{500}$ (see, for instance, the Tables II and III from the Ref. \cite{2013MNRAS.431.1487P}). Therefore, our results are in tension with these results from cosmological simulations. However, it is important to comment that the  models describing the physics of the intra-cluster medium used in hydrodynamic simulations may not span the entire range of physical process allowed by our current understanding.

\begin{figure*}[t]
    \centering
    \includegraphics[width=9.8cm, height=6.8cm]{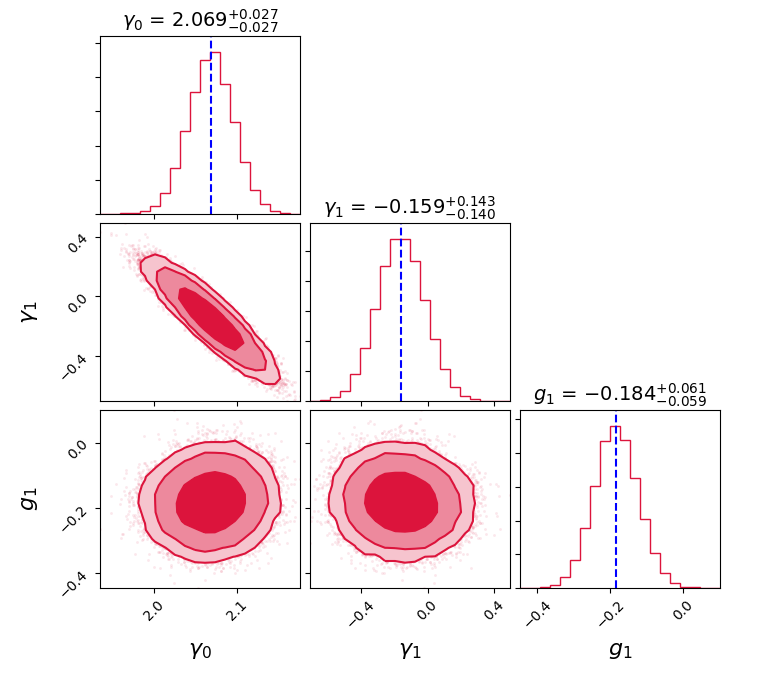} 
    \caption{Constraints on $\gamma_0$, $\gamma_1$, and $g_1$ parameters for the Intermediate mass range SGL sample. Here, $\gamma_1$ encodes the redshift evolution of $\gamma$. We find only a marginal 1.1$\sigma$ decrease with redshift for $\gamma$.   }
    \label{fig:interandgamma}
\end{figure*}

Finally, as we can see, the high SGL sub-sample is in full agreement with the SIS model ($\gamma=2$) (cf. Table~\ref{tab:table1}) while the low and intermediate mass SGL sub-samples are not compatible with this model  even at 3$\sigma$ c.l.. Therefore, our results also reinforce the need for  segregating the  lenses into low, intermediate, and high velocity dispersions, and analyzing them separately. 
As we obtain a mild evolution for the  gas depletion factor for Intermediate Mass Sample,  as a sanity check, we also look for the redshift dependence of the parameter $\gamma$ for the Intermediate mass sample. For this purpose, we posit a parametric form: $\gamma(z) = \gamma_0 + \gamma_1 z_l$ and set $\gamma_0$ and $\gamma_1$ as the free parameters along with gas depletion factor $g_1$. We get: $\gamma_0 = 2.069\pm0.027$, $\gamma_1 = -0.159 \pm 0.142$ and $g_1 = -0.184\pm0.06$ as shown in Figure~\ref{fig:interandgamma}. We found a marginal decrease  for $\gamma$ with redshift at 1.1$\sigma$. As we can see, we still get a mild evolution of gas depletion factor at  3.1$\sigma$.

\section{Conclusions}
\label{sec:conclusions}

In this letter, we have explored a possible  time evolution of the gas depletion factor, using a sample of 40 galaxy clusters with their X-ray gas mass fraction obtained in spherical shells at radii near $r_{2500}$. The analyses were performed by using the  gas mass fraction measurements in conjunction with  sub-samples of strong gravitational lens systems and the cosmic distance duality relation validity. The depletion factor was modelled assuming $g(z)=g_0 (1 + g_1 z)$, and we found a mild evolution at 2.7$\sigma$, i.e. $g_1 = -0.15 \pm 0.055$  (see Fig.~\ref{fig:evolution}). This estimate is obtained by  calculating an error-weighted average by combining  the different   values in Table ~\ref{tab:table1}, which contain the results for  each of the sub-samples.  

In contrast to previous work in this area, our method does not use the Planck constraints on  $\Omega_b/\Omega_M$, although we did assume a  flat universe.  This is the first work in literature which finds a non-negligible evolution of the gas depletion factor using gas mass fraction measurements in spherical shells at radii near $r_{2500}$. This value is in tension with the results from hydrodynamical simulations.
Moreover,  our results also reinforce the need for  segregating the  lenses into low, intermediate and high velocity dispersions, and analyzing them separately.  The  mass-sheet degeneracy in the gravitational lens system (see \cite{2016JCAP...08..020B} and references therein) and its effect on our results also could be explored as an interesting extension of this work. \rthis{We also note that our results are shown for interpolation carried out using GPR. Other reconstruction techniques could change the results somewhat.}

It is important to comment that the method discussed here can be used in a near future with data set   from the X-ray survey eROSITA \cite{eRosita}, that is expected to detect $\approx$ 100,000 galaxy clusters, along with followup optical and infrared data from EUCLID mission, Vera Rubin LSST, and Nancy Grace Rowan space telescope, that will discover thousands of strong lensing systems.  Then,  in the near future, as more and larger data sets with smaller statistical and systematic uncertainties become available, someone will can check the influence of cosmic curvature on our results, given that non-zero curvature has been found in some cosmological analyses~\cite{divalentino19}.

\section*{ACKNOWLEDGEMENTS}
KB would like to express his gratitude towards the Department of Science and Technology, Government of India for providing the financial support under DST-INSPIRE Fellowship program. RFLH
thanks CNPq No.428755/2018-6 and 305930/2017-6. \rthis{We are grateful to the anonymous referee for many constructive comments and feedback on our manuscript.}

    

    

\bibliography{ref}

\section*{Appendix}
\rthis{Here, we provide a brief introduction to GPR and also compare with other reconstruction techniques such as Artificial Neural Networks (ANN).}

{\it GPR:} 
\rthis{GPR is a non-parametric  method to interpolate between data points to reconstruct the original function at any input values. It has been widely used in over 200 papers in Cosmology. (See ~\cite{seikel12,Haveesh,BoraCDDR,LiuGPR} and references therein for a sampling of some of its applications to Cosmology.)  We provide an abridged description of GPR. More details can be found in some of the aforementioned works.}

\rthis{The Gaussian process is characterized by  a  mean function $\mu(x)$ and a covariance function $ cov(f(x),f(\Tilde{x})) = k(x,\Tilde{x}) $, which connects the values of $f$, when  evaluated at $x$ and $\Tilde{x}$. For a Gaussian kernel, $k(x,\Tilde{x})$  can be described by: $$ k(x,\Tilde{x}) = \sigma_f^2 \exp \left( -\frac{(x-\Tilde{x})^2}{2l^2} \right), $$ where  $\sigma_f$ and $l$ are the  hyperparameters describing  the `bumpiness' of the function. }

\rthis{For a set of input points, $\mathbf{X} \equiv x_i$, one can generate a vector $\mathbf{f^*}$ of function values evaluated at $\mathbf{X^*}$ with $f^*_i = f(x_i^*)$ as $$ \mathbf{f^*} = \mathcal{N}(\mathbf{\mu^*},K(\mathbf{X^*,X^*})) $$ 
Here, $\mathcal{N}$ implies  that  the Gaussian process is evaluated at $x^*$, where $f(x^*)$ is a random value drawn from a normal distribution. Similarly, observational data can be written in as $$ \mathbf{y} = \mathcal{N}(\mathbf{\mu},K(\mathbf{X,X})+C) $$ where $C$ is the covariance matrix of the data. For uncorrelated data, the covariance matrix is simply $diag(\sigma_i^2)$. Using the values of $y$ at $\mathbf{X}$, one  can reconstruct $\mathbf{f^*}$ using $$ \overline{\mathbf{f^*}} = \mathbf{\mu^*} + K(\mathbf{X^*,X})[K(\mathbf{X,X})+C]^{-1}(\mathbf{y-\mu}) $$ and $$ cov(\mathbf{f^*}) = K(\mathbf{X^*,X^*}) - K(\mathbf{X^*,X})[K(\mathbf{X,X})+C]^{-1}K(\mathbf{X,X^*}) $$ where $\overline{\mathbf{f^*}}$ and $cov(\mathbf{f^*})$ are mean and covariance of $\mathbf{f^*}$ respectively. The diagonal elements of $cov(\mathbf{f^*})$ provide us the variance of $\mathbf{f^*}$. Our implementation of GPR was implemented using the {\tt sklearn} package. }

\begin{figure*}
    \centering
    \includegraphics[width=10cm, height=8cm]{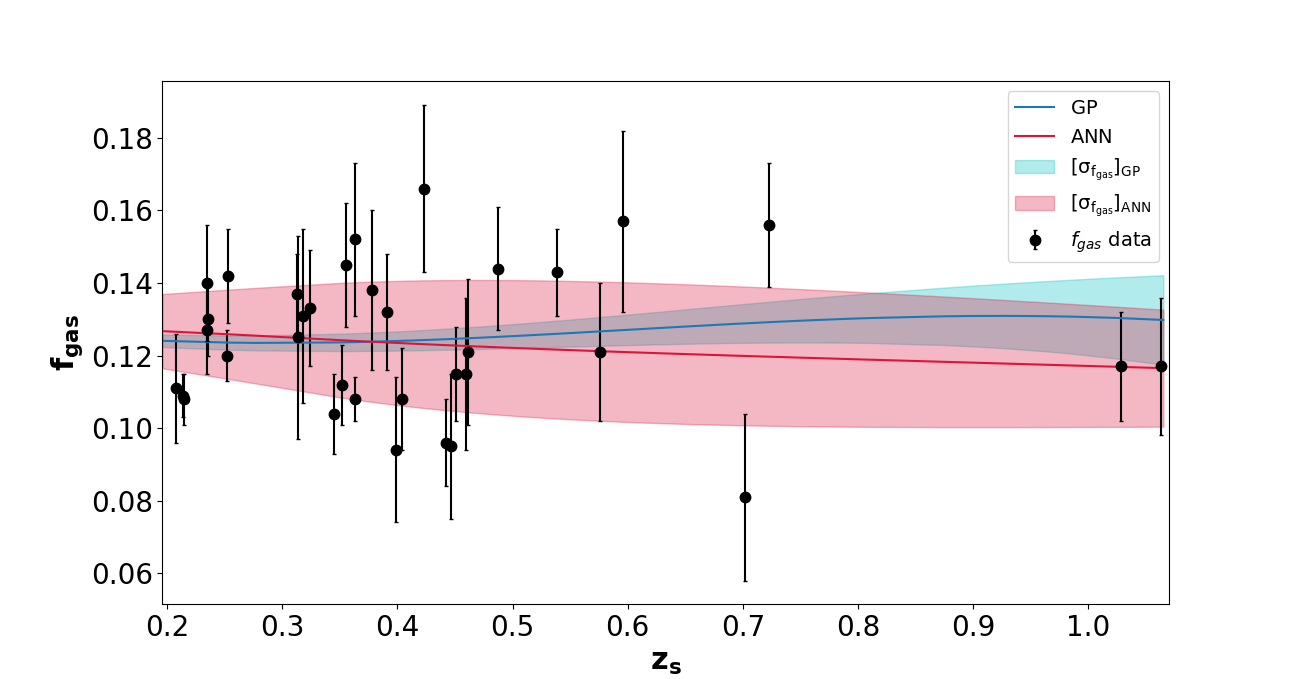}

    \caption{\rthis{A comparison between the Gaussian Processes reconstruction and Artificial Neural Networks (ANN) reconstruction of the gas mass fraction at $z_s$ for the Intermediate Mass range sample.  The shaded regions show the $1\sigma$ error bars from both the reconstruction techniques.}}
    \label{fig:ANN}
\end{figure*}

{\it ANN:} 
\rthis{ANN is another non-parameteric regression technique which has become popular following the increasing usage of Machine Learning applications to Astrophysics~\cite{Brunner,Bethapudi}.  The ANN consists of an input layer, multiple hidden layers and an output layer.  During each layer, a linear transformation is applied to the vector from the previous layer followed by a non-linear activation function, which is then propagated to the next layer. Different choices for the activation function and details of reconstruction using ANN is reviewed in ~\cite{LiuGPR}. For our analysis we use the publicly available code for ANN based regression in ~\cite{WangANN}.  A comparison of $f_{gas}$ reconstruction using both the techniques is illustrated in Fig.~\ref{fig:ANN}. We can see that there is a difference in the reconstructed $f_{gas}$ between the two methods. }

\end{document}